\documentclass[letterpaper]{article} 
\usepackage{aaai2027}  
\usepackage[hyphens]{url}  
\usepackage{graphicx} 
\usepackage{natbib}  
\usepackage{caption} 
\usepackage{algorithm}
\usepackage{algorithmic}

\usepackage{newfloat}
\usepackage{listings}
\DeclareCaptionStyle{ruled}{labelfont=normalfont,labelsep=colon,strut=off} 
\floatstyle{ruled}
\newfloat{listing}{tb}{lst}{}
\floatname{listing}{Listing}

\usepackage{booktabs}
\usepackage{algorithm}
\usepackage{algorithmic}
\usepackage{booktabs}
\usepackage{amsfonts}
\usepackage{amsmath}
\usepackage{adjustbox}
\usepackage{multirow}
\usepackage{makecell}
\usepackage{graphicx}

\usepackage{tabularx}
\usepackage{booktabs}
\usepackage{array}

\usepackage{amsthm}

\newcolumntype{C}{>{\centering\arraybackslash}X}

\usepackage[table]{xcolor}
\usepackage{booktabs}
\usepackage{multirow}
\usepackage{graphicx}

\definecolor{phasebg}{RGB}{250,246,232}
\definecolor{gaincolor}{RGB}{35,145,80}
\definecolor{losscolor}{RGB}{180,60,50}

\title{AnchorMark: Robust Diffusion Watermarking via Latent-Space Rotation Synchrony}

\author {
    Yuqi Qian\textsuperscript{\rm 1,\rm 2},
    Yun Cao\textsuperscript{\rm 1,\rm 2},
    Haocheng Fu\textsuperscript{\rm 1,\rm 2},
    Haochen Zhao\textsuperscript{\rm 1,\rm 2},
    Hong Zhang\textsuperscript{\rm 1,\rm 2},
    Meineng Zhu\textsuperscript{\rm 3},
}
\affiliations {
    \textsuperscript{\rm 1}Institute of Information Engineering, Chinese Academy of Sciences\\
    \textsuperscript{\rm 2}School of Cyber Security, University of Chinese Academy of Sciences\\
    \textsuperscript{\rm 3}School of Cybersecurity, University of International Relations\\
    qianyuqi@iie.ac.cn, caoyun@iie.ac.cn, fuhaocheng@iie.ac.cn, zhaohaochen@iie.ac.cn, zhanghong@iie.ac.cn, zmneng@uir.edu.cn 
}

\begin{document}

\maketitle

\begin{abstract}
Inversion-based watermarking embeds watermark payloads directly into the generative process, avoiding a separate post-hoc image-domain embedding stage while preserving the native visual fidelity of synthesized images.
However, existing methods remain vulnerable to compound lossy post-processing, particularly when rotation is involved, as it disrupts the spatial correspondence required for latent-space decoding. 
To overcome this limitation, we introduce \textbf{AnchorMark}, a training-free, robust inversion-based watermarking. We uncover a latent-space property termed \emph{Rotation Synchrony}: image-domain rotations and their counterparts in the recovered initial latent share the same angle. Building on this property, AnchorMark embeds a synchronization anchor in the central region of the initial latent, enabling accurate estimation and correction of the rotation angle during extraction.
Experiments show that AnchorMark substantially improves bit accuracy under rotation and combined attacks, with limited impact on image quality.

\end{abstract}

\section{Introduction}

\begin{figure}[t]
  \centering
  \includegraphics[width=\columnwidth,
    trim=2cm 2cm 0cm 2cm,
    clip
  ]{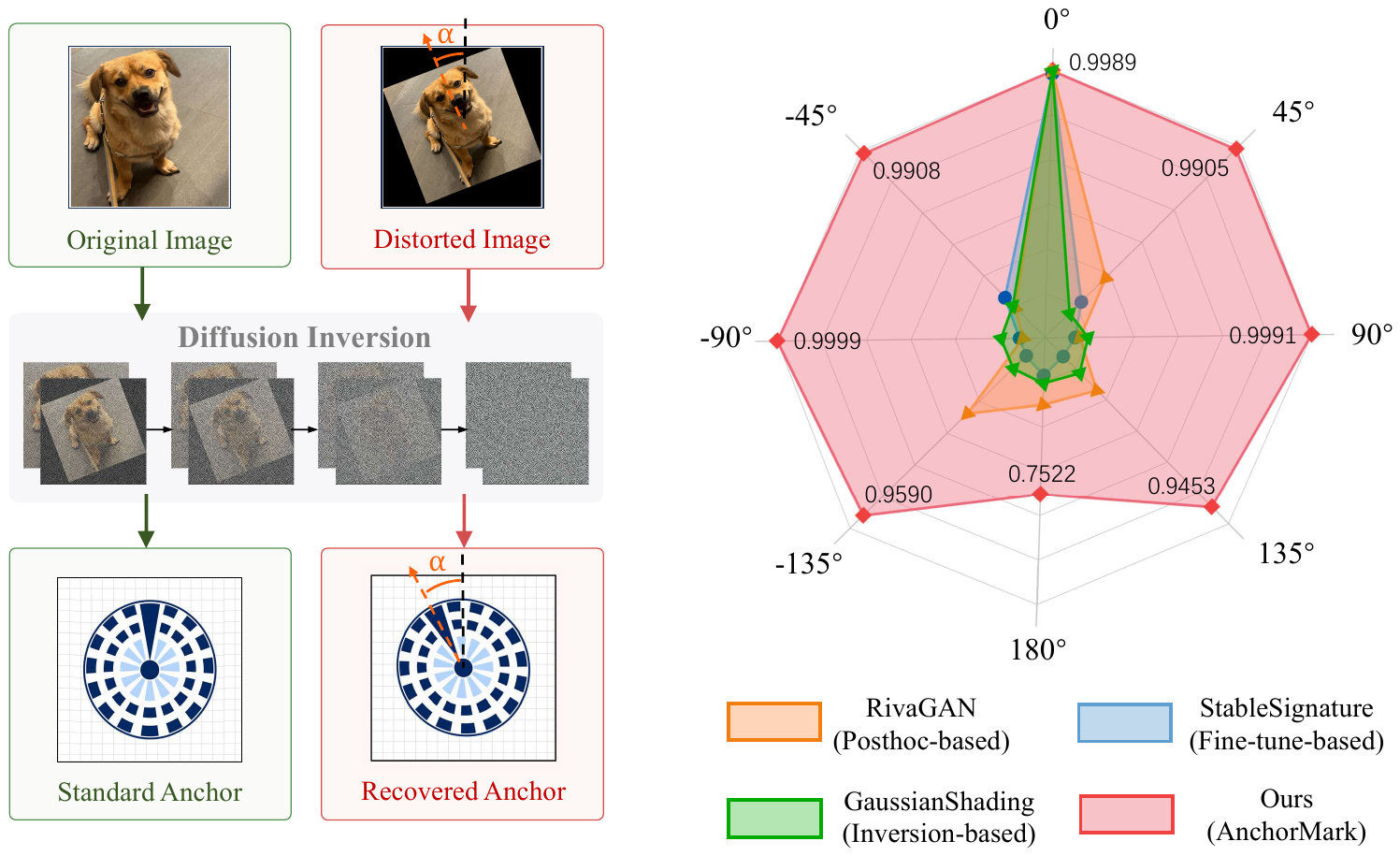}
\caption{Intuitive comparison of AnchorMark with representative state-of-the-art methods from three watermarking paradigms. AnchorMark maintains high payload extraction accuracy across a wide range of rotation angles.}
  \label{fig:example}
\end{figure}

Recent advances in diffusion models have substantially improved the quality and controllability of text-to-image generation
\cite{ho2020denoising,song2020denoising,rombach2022high}.
Latent diffusion models such as Stable Diffusion can now synthesize realistic, semantically rich, and stylistically diverse images from natural-language prompts, while also raising concerns about content provenance, copyright protection, model-output authentication, and synthetic misinformation.
Digital watermarking has become a crucial technique to address this need.
Among existing paradigms, inversion-based watermarking is particularly appealing: it embeds payloads into the initial diffusion latent, allowing the watermark to be formed together with the image, and later recovers it through diffusion inversion.
This formulation requires no retraining of the underlying generator and is robust to common non-geometric distortions
\cite{wen2023tree,yang2024gaussian,qian2026shapemark}.
Nevertheless, their robustness deteriorates markedly under combinations of lossy post-processing operations, particularly when rotation is involved.
As shown in Fig.~\ref{fig:example}, under rotations spanning $-180^\circ$ to $180^\circ$, the bit accuracy of representative watermarking methods drops toward $50\%$ over a broad range of angles, indicating nearly random payload recovery rather than a gradual loss of watermark strength.

The fundamental challenge is latent coordinate-frame misalignment: rotation shifts watermark carriers from their expected locations, breaking the spatial correspondence required for payload decoding. Prior work responds to rotation-induced misalignment in two different ways. Learned synchronizers explicitly estimate the applied transformation using auxiliary networks or predictors
\cite{fang2025syntag}, but require additional training and can make synchronization dependent on particular architectures, watermarking schemes, or training attack distributions. Rotation-invariant spectral designs instead avoid explicit transformation estimation by suppressing orientation sensitivity
\cite{wen2023tree,ci2024ringid}. Although effective for watermark-presence detection, such designs do not recover the correspondence between latent carriers and bit positions required to decode an ordered multi-bit payload. 
Inversion-based multi-bit watermarking therefore still lacks a lightweight and modular mechanism for watermarking synchronization.

To address this challenge, we introduce \textbf{AnchorMark} as a training-free robust inversion-based watermarking. 
Inspired by the classical rotation property of the Fourier transform, which maps an image-space rotation to the same angular rotation in the frequency domain, we investigate how rotation propagates through diffusion latent recovery.
We uncover \emph{Rotation Synchrony} in diffusion latent recovery: an image-space rotation induces the same angular rotation in the recovered initial latent. Consequently, matching the recovered latent against rotated references yields a minimum near the ground-truth angle.
Based on this property, AnchorMark embeds a compact synchronization anchor in the central region of the initial latent while retaining the surrounding region for payload embedding. During recovery, the anchor guides rotation estimation and refinement, after which the aligned image is passed to the original decoder for payload extraction.
The main contributions are summarized as follows:

\begin{itemize}
    \item We identify rotation-induced latent coordinate misalignment as a fundamental failure mode and formalize Rotation Synchrony: an image-space rotation induces the same angular rotation in the recovered initial latent.

    \item We propose \textbf{AnchorMark}, a training-free robust inversion-based watermarking. It combines a compact multi-frequency phase anchor with statistics-calibrated injection, coarse-to-fine registration, and local refinement guided by payload verification to restore bit-wise alignment.

    \item Extensive experiments demonstrate that AnchorMark consistently improves robustness against rotation attacks and enables reliable multi-bit payload recovery while preserving the visual quality and semantic fidelity of generated images.
\end{itemize}

\section{Related Work}

Digital watermarking embeds traceable information into visual content while preserving perceptual quality. Classical transform-domain and neural post-hoc methods modify images after synthesis
\cite{ingemar2008digital,zhu2018hidden,zhang2019robust,tancik2020stegastamp,fernandez2022watermarking},
whereas recent generative watermarking methods integrate the signal into the image-generation process. Stable Signature fine-tunes the latent decoder to produce signed images
\cite{fernandez2023stable}, while Tree-Ring, Gaussian Shading, and ShapeMark embed watermarks into the initial diffusion latent and recover them through inversion
\cite{wen2023tree,yang2024gaussian,qian2026shapemark}. CoSDA further improves this paradigm by reducing inversion drift and compensation errors
\cite{fang2025cosda}. Although effective against common non-geometric distortions, inversion-based multi-bit methods associate payload bits with spatially indexed latent carriers and are therefore vulnerable to rotation-induced misalignment.

Geometric robustness has traditionally been addressed through synchronization references, feature-based registration, or invariant descriptors such as Fourier--Mellin features
\cite{bas2002geometrically}. Recent learned methods explicitly predict geometric transformations
\cite{fang2025syntag}, but require additional training and may depend on specific architectures or attack distributions. Rotation-invariant spectral designs, such as Tree-Ring and RingID
\cite{wen2023tree,ci2024ringid}, instead suppress orientation sensitivity to support stable watermark detection. However, multi-bit tracing requires recovering the correspondence between latent carriers and bit positions, which rotation-invariant aggregation does not provide. AnchorMark addresses this gap by preserving the spatially indexed latent structure and explicitly restoring its coordinate alignment for payload recovery.

\section{Rotation Synchrony in Latent Space}
\label{sec:rotation_synchrony}

AnchorMark is grounded in a structural property of latent space that we term \emph{Rotation Synchrony}: \textbf{an image-space rotation induces the same angular rotation in the recovered initial latent. } Prior work has shown that geometric transformations can remain systematically organized in learned representations \cite{lenc2015understanding}. Although standard CNNs are not strictly rotation-equivariant \cite{worrall2017harmonic}, transformation-dependent spatial structures may persist in their feature spaces \cite{cohen2016group,kouzelis2025eq}.

\subsection{Residual-Bounded Rotation Synchrony}

Let \(x_T\) denote the initial diffusion latent, and let
\(z_0=\Phi(x_T)\) be the final latent produced by the
deterministic sampling map \(\Phi\).
Let \(\mathcal{E}\) and \(\mathcal{D}\) denote the VAE encoder
and decoder, respectively, and let \(\Psi\) denote deterministic
diffusion inversion. The generated image is
\(I=\mathcal{D}(z_0)\). After applying an image-domain rotation
\(\mathcal{T}_{\alpha}\), the recovered initial latent is
\begin{equation}
    \hat{x}_T(\alpha)
    =
    \Psi\!\left(
        \mathcal{E}\!\left(
            \mathcal{T}_{\alpha}\mathcal{D}(z_0)
        \right)
    \right).
    \label{eq:rotated_image_inversion}
\end{equation}
Throughout the analysis, \(\mathcal{R}_{\alpha}\) denotes the
corresponding norm-preserving rotation on the continuous latent
grid. Deviations introduced by finite-grid interpolation and finite
spatial support are absorbed into the residual terms below.

\noindent
\textbf{Definition 1 (Rotation Synchrony).}
Given a generated image--latent pair \((I,x_T)\), the diffusion
generation--inversion pipeline exhibits \(B_{\alpha}\)-bounded
\emph{Rotation Synchrony} at angle \(\alpha\) if
\begin{equation}
    \left\|
        \hat{x}_T(\alpha)
        -
        \mathcal{R}_{\alpha}x_T
    \right\|_2
    \leq B_{\alpha}.
    \label{eq:rotation_synchrony_definition}
\end{equation}
Rotation Synchrony therefore characterizes an approximate
cross-domain commutativity relation: rotating a generated image
before latent recovery approximately agrees with directly rotating
its initial latent. Importantly, this property does not assume exact
rotation equivariance of either the VAE or the diffusion inversion
dynamics.

We first characterize the deviation introduced by VAE encoding.
Its response to image rotation admits the decomposition
\begin{equation}
    \mathcal{E}\!\left(
        \mathcal{T}_{\alpha}\mathcal{D}(z_0)
    \right)
    =
    \mathcal{R}_{\alpha}z_0
    +
    r_{\mathrm{vae}}(\alpha,z_0),
    \label{eq:vae_rotation_decomposition}
\end{equation}
where \(r_{\mathrm{vae}}\) quantifies the VAE-mediated deviation
from ideal Rotation Synchrony, including reconstruction
inconsistency, finite-resolution interpolation, and boundary
effects. Equation~\eqref{eq:vae_rotation_decomposition} is an exact
residual decomposition and does not assume that the VAE is
rotation-equivariant.

We next characterize how this rotational relation propagates
through deterministic diffusion inversion. Following the ODE
interpretation of deterministic diffusion trajectories
\cite{song2020denoising,song2020score}, we write the inversion
dynamics as
\begin{equation}
    \frac{dx(t)}{dt}
    =
    v_t(x(t)),
    \qquad
    x(0)=z_0,
    \qquad
    x(T)=\Psi(z_0).
\end{equation}
We define the rotation-compatibility defect of the inversion vector
field as
\begin{equation}
    \delta_t(\alpha,x)
    =
    v_t(\mathcal{R}_{\alpha}x)
    -
    \mathcal{R}_{\alpha}v_t(x).
    \label{eq:rotation_compatibility_defect}
\end{equation}
If the inversion dynamics were exactly rotation-compatible, then
\(\delta_t(\alpha,x)=0\). For a practical diffusion model, this
quantity measures the discrepancy between rotating a latent before
inversion and rotating the corresponding inversion trajectory.

Assume that \(v_t\) is locally Lipschitz along the relevant
trajectories, with Lipschitz constant \(L_t\). A standard Gronwall
argument gives
\begin{equation}
\begin{aligned}
&
\left\|
    \Psi(\mathcal{R}_{\alpha}z_0)
    -
    \mathcal{R}_{\alpha}\Psi(z_0)
\right\|_2
\\
&\quad\leq
\int_0^T
\exp\!\left(
    \int_s^T L_{\tau}\,d\tau
\right)
\left\|
    \delta_s(\alpha,x(s))
\right\|_2
\,ds .
\end{aligned}
\label{eq:inversion_rotation_bound}
\end{equation}
Thus, the inversion stage preserves the rotational relation required
by Rotation Synchrony up to the accumulated
rotation-compatibility defect of its vector field. The above
analysis is stated for the continuous deterministic inversion flow;
the corresponding finite-step DDIM formulation and its
discretization residual are provided in the appendix.

\begin{figure}[t]
  \centering
  \includegraphics[width=\columnwidth,
        trim=1cm 0cm 0cm 1cm,
        clip]{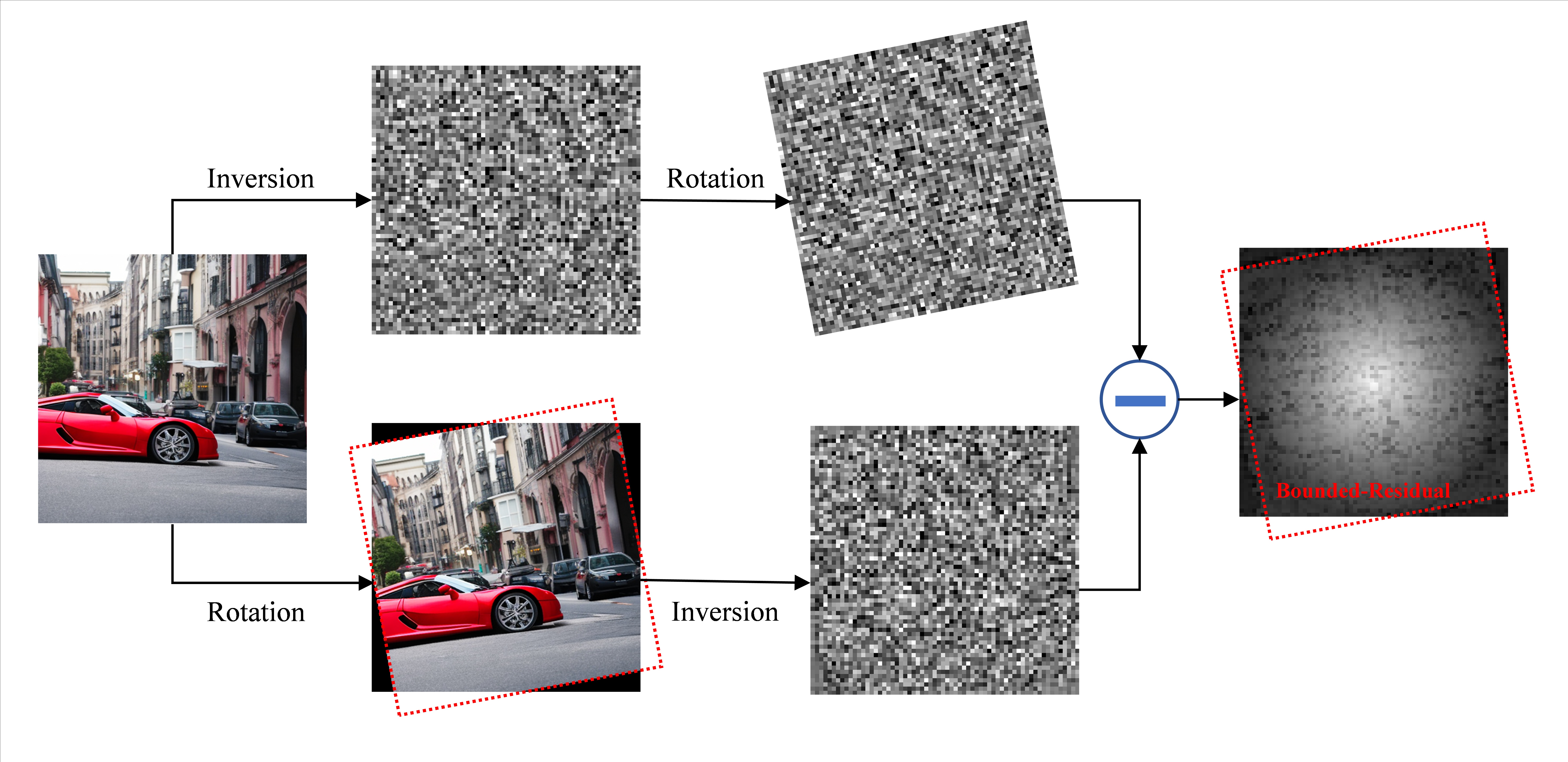}
  \caption{Illustration of Rotation Synchrony: the initial latent recovered from a rotated image closely matches the correspondingly rotated initial latent of the original image.}
  \label{difference}
\end{figure}

If the inversion map is locally Lipschitz in a neighborhood of
\(\mathcal{R}_{\alpha}z_0\), with constant \(L_{\Psi}\), the VAE
residual propagates according to
\begin{equation}
\begin{aligned}
&
\left\|
    \Psi\!\left(
        \mathcal{R}_{\alpha}z_0
        +
        r_{\mathrm{vae}}
    \right)
    -
    \Psi(\mathcal{R}_{\alpha}z_0)
\right\|_2
\\
&\qquad\leq
L_{\Psi}
\left\|
    r_{\mathrm{vae}}(\alpha,z_0)
\right\|_2 .
\end{aligned}
\label{eq:vae_residual_propagation}
\end{equation}

Combining Eqs.~\eqref{eq:inversion_rotation_bound} and
\eqref{eq:vae_residual_propagation} yields the following
residual-bounded characterization of Rotation Synchrony.

\noindent
\textbf{Proposition 1 (Residual-Bounded Rotation Synchrony).}
Under the above local Lipschitz conditions, for a generated image
\(I=\mathcal{D}(\Phi(x_T))\) rotated by \(\alpha\), the
inversion-recovered initial latent admits the decomposition
\begin{equation}
    \hat{x}_T(\alpha)
    =
    \mathcal{R}_{\alpha}x_T
    +
    \xi_{\alpha},
    \label{eq:rotation_propagation_model}
\end{equation}
where
\begin{equation}
\begin{aligned}
\left\|\xi_{\alpha}\right\|_2
&\leq
L_{\Psi}
\left\|
    r_{\mathrm{vae}}(\alpha,z_0)
\right\|_2
\\
&\quad+
\int_0^T
\exp\!\left(
    \int_s^T L_{\tau}\,d\tau
\right)
\left\|
    \delta_s(\alpha,x(s))
\right\|_2
\,ds
\\
&\quad+
\left\|
    \Psi(\Phi(x_T))-x_T
\right\|_2 .
\end{aligned}
\label{eq:combined_rotation_residual}
\end{equation}

The three terms respectively quantify the VAE-mediated rotation
residual, the accumulated rotation-compatibility defect of the
inversion dynamics, and the generation--inversion cycle error.
Together, they characterize the deviation from ideal Rotation
Synchrony. If all three terms vanish,
Eq.~\eqref{eq:rotation_propagation_model} reduces to
\(\hat{x}_T(\alpha)=\mathcal{R}_{\alpha}x_T\), and the pipeline is
exactly rotation-synchronous. In practical pipelines,
Eq.~\eqref{eq:combined_rotation_residual} establishes the
residual-bounded form of this property, with its right-hand side
providing an explicit choice of \(B_{\alpha}\) in
Definition~1. The practical magnitude of these residuals, and
whether they remain sufficiently small to support angle recovery,
are examined empirically below.

\subsection{Rotation Estimation via Latent Matching}

The above result motivates rotation estimation through latent matching.
For a candidate angle \(\beta\), let
\begin{equation}
y_{\alpha}
=
\operatorname{Norm}\!\left(\hat{x}_T(\alpha)\right),
\qquad
z_{\beta}
=
\operatorname{Norm}\!\left(\mathcal{R}_{\beta}x_T\right).
\end{equation}
We define the practical and ideal matching distances as
\begin{align}
D_{\mathrm{rec}}(\alpha,\beta)
&=
\frac{1}{N}
\left\|
y_{\alpha}-z_{\beta}
\right\|_2^2,
\label{eq:recovered_matching_distance}
\\
D_{\mathrm{ideal}}(\alpha,\beta)
&=
\frac{1}{N}
\left\|
z_{\alpha}-z_{\beta}
\right\|_2^2.
\label{eq:ideal_matching_distance}
\end{align}
Without recovery residuals, \(y_{\alpha}=z_{\alpha}\), and thus
\(D_{\mathrm{rec}}=D_{\mathrm{ideal}}\).
For a reference latent without nontrivial rotational symmetry,
\(D_{\mathrm{ideal}}(\alpha,\beta)\) is minimized at
\(\beta=\alpha\).

Let
\begin{equation}
\epsilon_{\alpha}
=
\frac{1}{\sqrt{N}}
\left\|
y_{\alpha}-z_{\alpha}
\right\|_2
\end{equation}
denote the normalized recovery deviation. By the reverse triangle inequality,
\begin{equation}
\left|
\sqrt{D_{\mathrm{rec}}(\alpha,\beta)}
-
\sqrt{D_{\mathrm{ideal}}(\alpha,\beta)}
\right|
\leq
\epsilon_{\alpha}.
\label{eq:matching_profile_perturbation}
\end{equation}
Hence, the practical matching profile is a bounded perturbation of its ideal counterpart. When the VAE and inversion residuals are small, its minimum is expected to remain near the ground-truth rotation, which we verify empirically below.

\begin{figure}[t]
  \centering
  \includegraphics[width=\columnwidth,
        trim=100 0 0 100,
        clip]{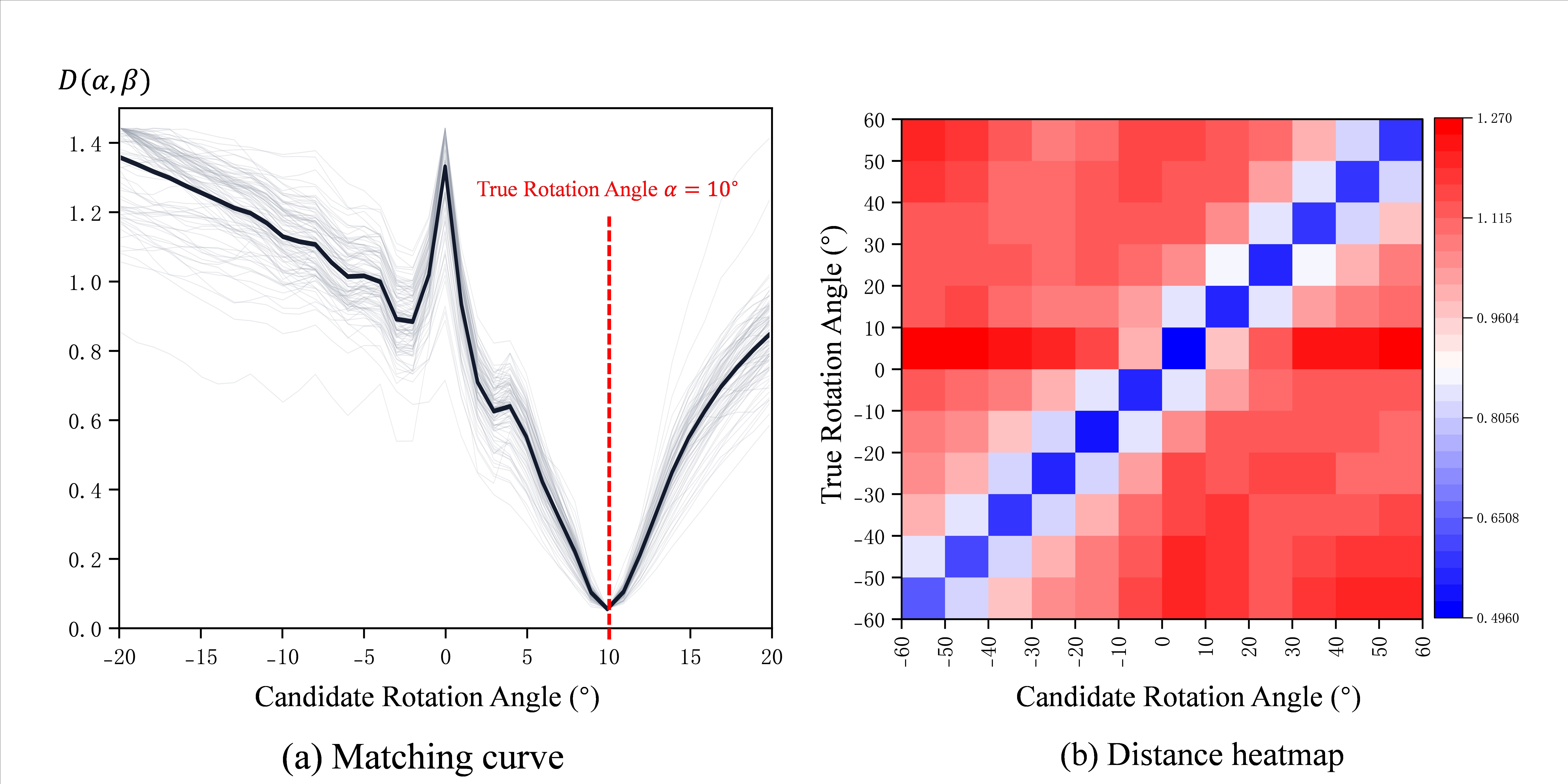}
  \caption{Empirical validation of Rotation Synchrony: latent matching is minimized near the ground-truth angle, producing a clear low-distance diagonal across rotations.}
  \label{heatmap}
\end{figure}

\subsection{Empirical Validation of Rotation Synchrony}

We empirically assess whether Rotation Synchrony is sufficiently accurate to support angle recovery in pretrained diffusion pipelines. For each initial latent \(x_T\), we generate
\(I=\mathcal{D}(\Phi(x_T))\), rotate it by a known angle \(\alpha\), and apply DDIM inversion to obtain \(\hat{x}_T(\alpha)\). We then evaluate
\(D_{\mathrm{rec}}(\alpha,\beta)\) against candidate rotations
\(\mathcal{R}_{\beta}x_T\), using the central valid region to reduce rotation-induced boundary artifacts.

As shown in Fig.~\ref{heatmap}, both the sample-wise and averaged matching profiles attain their minima at the ground-truth angle or its nearest sampled candidate. The pairwise distance heatmap further exhibits a clear low-distance diagonal, providing a global signature of Rotation Synchrony. Consequently,
\begin{equation}
    \hat{\alpha}
    =
    \arg\min_{\beta}
    D_{\mathrm{rec}}(\alpha,\beta)
    \approx
    \alpha.
\end{equation}
These results show that image rotation remains observable as the dominant orientation shift of the inversion-recovered initial latent, providing the empirical basis for AnchorMark to perform latent synchronization.

\begin{figure*}[t]
    \centering
    \includegraphics[
        width=\textwidth,
        trim=0cm 3cm 0cm 3cm,
        clip
    ]{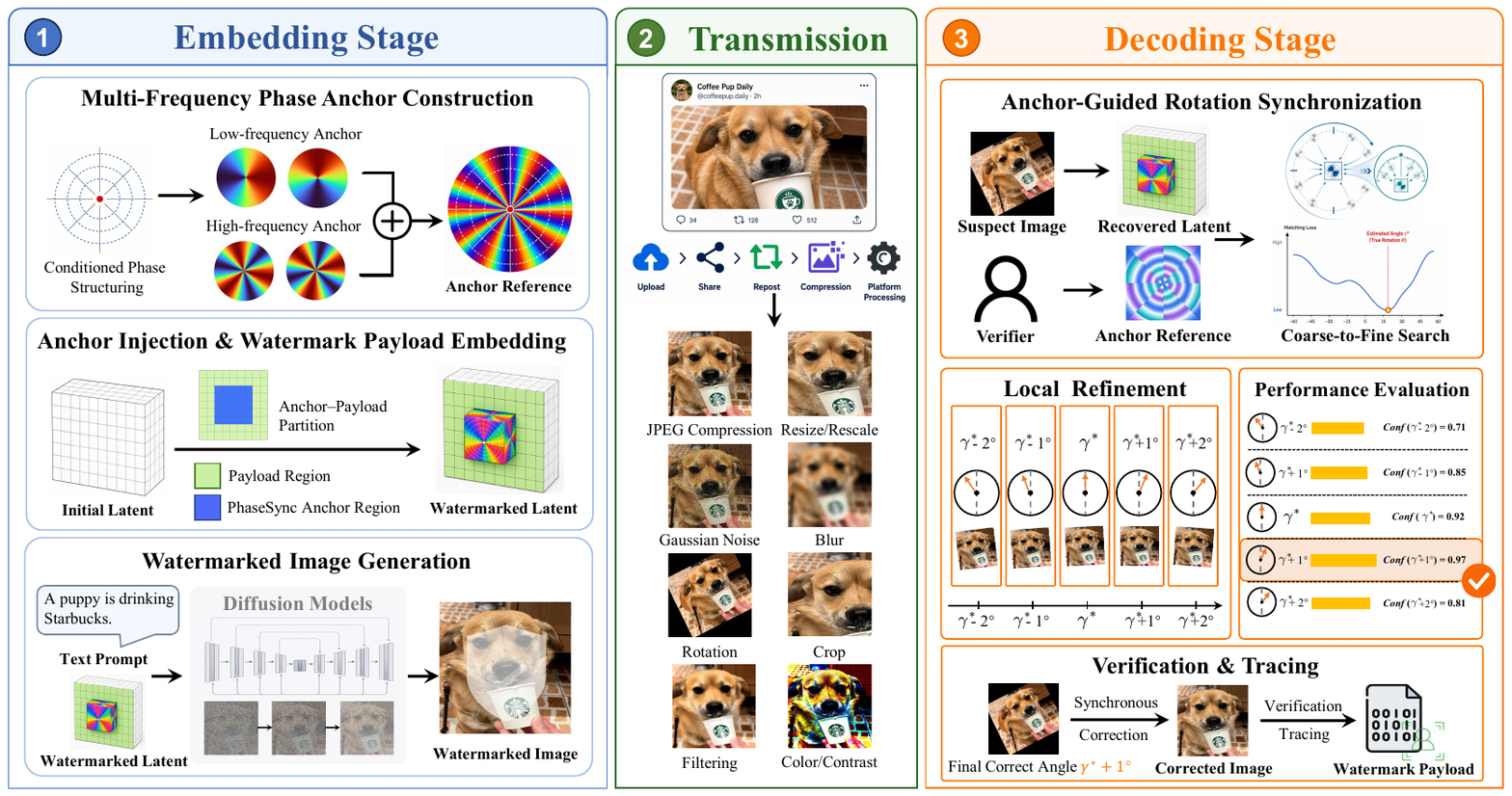}
    \caption{Overall pipeline of AnchorMark. A multi-frequency phase anchor is injected into the initial diffusion latent together with the watermark payload; after transmission under common distortions, the recovered anchor guides coarse-to-fine rotation synchronization and local refinement for reliable payload verification and tracing.}
    \label{fig:framework}
\end{figure*}

\section{AnchorMark: Rotation Synchronization for Inversion-Based Watermarking}
In this section, we present \textbf{AnchorMark}, a training-free geometric synchronization framework for inversion-based diffusion watermark recovery. Building on the rotation correspondence established above, AnchorMark embeds a compact multi-frequency phase anchor in the central latent region through statistics-calibrated injection, while retaining the surrounding region for the original multi-bit payload. As illustrated in Fig.~\ref{fig:framework}, recovery combines anchor-guided coarse-to-fine rotation synchronization with local refinement, reusing the original watermark decoder for candidate verification and final payload extraction after image-domain correction.

\subsection{Decoupled Phase-Anchor Construction}
\label{sec:phase_anchor}

AnchorMark constructs an analytic orientation reference that is spatially decoupled from the data-bearing watermark payload. The construction follows three principles: isolating the synchronization cue from payload modulation, encoding rotation through a predictable multi-frequency phase response, and limiting perturbations to the initial latent through statistics-calibrated injection.

\paragraph{Anchor--Payload Spatial Decoupling.}
Given an initial diffusion latent
\(x_T\in\mathbb{R}^{C\times H\times W}\),
AnchorMark reserves a compact central patch \(M_a\) for synchronization and a channel subset
\(\mathcal{S}_a\subseteq\{1,\ldots,C\}\)
for carrying the phase anchor. The resulting anchor support is therefore defined jointly by the central spatial region and the selected latent channels. Phase-anchor injection is restricted to this support, while the underlying inversion-based watermarking method embeds its multi-bit payload in the remaining latent coordinates.

This separation assigns synchronization and message carrying to disjoint latent supports, preventing payload modulation from overwriting the orientation reference and avoiding direct interference from the anchor with payload carriers. The anchor is placed at the latent center because centered image rotations preserve the central structure more consistently, whereas peripheral regions are more susceptible to padding, cropping, and interpolation artifacts. The central support therefore provides a stable and modular reference for subsequent rotation registration.

\paragraph{Multi-Frequency Phase Encoding.}
Within the \(s\times s\) anchor patch, each spatial position \((u,v)\) is represented in polar coordinates relative to the patch center \((u_c,v_c)\):
\begin{equation}
\begin{aligned}
\theta(u,v)
&= \operatorname{atan2}(v-v_c,\,u-u_c),\\
\rho(u,v)
&= \sqrt{(u-u_c)^2+(v-v_c)^2}.
\end{aligned}
\end{equation}

Let \(C_a=|\mathcal{S}_a|\) and
\begin{equation}
    P=\left\lfloor\frac{C_a}{2}\right\rfloor
\end{equation}
be the number of available sine--cosine channel pairs. We select an ordered angular-frequency set
\begin{equation}
    \mathcal{K}=\{k_j\}_{j=1}^{P},
    \qquad
    1=k_1<k_2<\cdots<k_P\leq k_{\max},
\end{equation}
where \(k_{\max}\) is limited by the anchor resolution to avoid severe angular aliasing. The \(j\)-th harmonic pair is defined as
\begin{equation}
    \mathbf{A}_j(u,v)
    =
    \omega(\rho(u,v))
    \begin{bmatrix}
    \cos\!\bigl(k_j\theta(u,v)+\phi_j\bigr)\\
    \sin\!\bigl(k_j\theta(u,v)+\phi_j\bigr)
    \end{bmatrix},
\end{equation}
where \(\phi_j\) is a predetermined phase offset and
\(\omega(\rho)\) is a radial envelope, with
\(\omega(\rho)=1\) when no radial tapering is applied. The two components of \(\mathbf{A}_j\) are assigned to one latent-channel pair.

This construction converts spatial rotation into an analytic phase response. Under a centered rotation by \(\alpha\), the angular coordinate changes from \(\theta\) to \(\theta-\alpha\), yielding
\begin{equation}
    \mathbf{A}_j(\theta-\alpha)
    =
    \mathbf{Q}(k_j\alpha)\mathbf{A}_j(\theta),
\end{equation}
where
\begin{equation}
\mathbf{Q}(k_j\alpha)
=
\begin{bmatrix}
\cos(k_j\alpha) & \sin(k_j\alpha)\\
-\sin(k_j\alpha) & \cos(k_j\alpha)
\end{bmatrix}.
\end{equation}
Thus, a rotation by \(\alpha\) induces a predictable phase displacement of \(k_j\alpha\) in the \(j\)-th harmonic pair. The fundamental component \(k_1=1\) provides a stable global orientation reference with limited angular ambiguity, whereas higher-frequency components amplify small angular variations and sharpen local discrimination. Their combination forms a multi-scale phase code that supports both reliable coarse alignment and fine-grained rotation estimation.

\paragraph{Statistics-Calibrated Anchor Injection.}
Directly overwriting the initial latent with a deterministic phase template may introduce a substantial distribution shift and degrade image generation. AnchorMark instead embeds the anchor through per-channel moment matching, weak mixing, and post-mixing calibration.

Let
\(x_a=x_T[M_a]\)
denote the original latent patch, and let \(A\) denote the corresponding multi-frequency phase template. For each selected channel \(c\), the template is first standardized and matched to the mean and standard deviation of the original latent patch:
\begin{equation}
\tilde{A}_c
=
\frac{A_c-\mu(A_c)}
     {\sigma(A_c)+\epsilon},
\qquad
A_c^{*}
=
\sigma(x_{a,c})\tilde{A}_c+\mu(x_{a,c}),
\end{equation}
where \(\mu(\cdot)\) and \(\sigma(\cdot)\) denote the empirical mean and standard deviation.

The matched template is then fused with the original latent through
\begin{equation}
x'_{a,c}
=
\sqrt{1-\lambda^2}\,x_{a,c}
+
\lambda A_c^{*},
\end{equation}
where \(\lambda\) controls the strength of the synchronization cue. A smaller \(\lambda\) reduces disturbance to the original latent, whereas a larger value strengthens the recoverable phase response. Finally, the mixed patch is recalibrated to the original per-channel moments:
\begin{equation}
x''_{a,c}
=
\sigma(x_{a,c})
\frac{x'_{a,c}-\mu(x'_{a,c})}
     {\sigma(x'_{a,c})+\epsilon}
+
\mu(x_{a,c}).
\end{equation}

The calibrated patch \(x''_a\) is written back to the anchor support of \(x_T\). This procedure approximately restores the original per-channel mean and variance up to the numerical stabilizer while retaining the structured phase component required for synchronization. Because the phase code and its injection are fully analytic, the anchor can be constructed without learned parameters or modifications to the diffusion model.

\subsection{Hierarchical Rotation Synchronization and Payload Recovery}

AnchorMark restores the latent coordinate frame through a hierarchical recovery procedure that combines two complementary signals. The phase anchor supports efficient rotation estimation over a wide angular range, whereas the original watermark decoder provides a task-specific criterion for resolving the remaining fine-grained uncertainty. Accordingly, AnchorMark first performs phase-anchor-guided global registration and then conducts payload-confidence-guided local refinement before final message recovery.

\paragraph{Anchor-Guided Global Registration.}
Given an attacked image \(I_{\mathrm{att}}\), AnchorMark first maps it back to the initial diffusion latent space:$\hat{x}_T=\mathrm{Inv}(I_{\mathrm{att}})$, and extracts the recovered anchor region $\hat{x}_a=\hat{x}_T[M_a]$. Using the shared key, the detector reconstructs the reference anchor \(x_a^{\mathrm{ref}}\). For each candidate attack angle \(\gamma\in\Omega\), the reference anchor is spatially rotated in the latent domain and compared with the recovered anchor through the normalized matching loss
\begin{equation}
\mathcal{L}_{\mathrm{anchor}}(\gamma)
=
\mathrm{MSE}
\left(
\mathrm{Norm}(\hat{x}_a),
\mathrm{Norm}\bigl(R_{\gamma}(x_a^{\mathrm{ref}})\bigr)
\right),
\end{equation}
where \(R_{\gamma}(\cdot)\) denotes latent-domain rotation. The initial rotation estimate is obtained as $
\gamma^{*}
=
\arg\min_{\gamma\in\Omega}
\mathcal{L}_{\mathrm{anchor}}(\gamma).
$
To cover a broad rotation range without exhaustively evaluating a dense angle grid, AnchorMark searches \(\Omega\) in a coarse-to-fine manner. Because this stage only rotates and compares compact latent anchors after a single inversion, it provides an efficient global estimate of the rotation-induced coordinate misalignment.

\paragraph{Payload-Guided Local Refinement.}
Although anchor-guided registration estimates the dominant rotation, small errors may remain because of finite search resolution, image interpolation, and diffusion inversion noise. Such residual misalignment can still disrupt the spatial correspondence required for multi-bit decoding. AnchorMark therefore constructs a compact local candidate set \(\Omega_v\) around the initial estimate \(\gamma^{*}\) and refines it using the native verification criterion of the underlying watermarking method.

For each \(\gamma\in\Omega_v\), the attacked image is inversely rotated by \(-\gamma\), mapped back to the initial latent space, and processed by the original payload decoder. Because AnchorMark confines synchronization to the central anchor region, the surrounding payload support and its native decoding rule remain unchanged. The payload recovered from this surrounding region is assigned the corresponding verification confidence \(\mathrm{Conf}(\gamma)\), and the final synchronization angle is selected as
\[
\hat{\alpha}
=
\arg\max_{\gamma\in\Omega_v}
\mathrm{Conf}(\gamma).
\]
The payload decoded at \(\hat{\alpha}\) is returned as the final recovery result. This hierarchical design uses the central phase anchor for wide-range synchronization and the surrounding native payload for task-aligned local refinement.

\begin{table*}[ht]
\centering
\caption{
Main comparison under rotation attacks. For each rotation magnitude, we report TPR / Acc, where TPR is evaluated at $\mathrm{FPR}=10^{-6}$ and Acc denotes bit-level tracing accuracy. ``--'' indicates not applicable. FID and CLIP-Score are reported as mean $\pm$ standard deviation.
}
\label{tab:main_results}
\renewcommand{\arraystretch}{1.12}
\setlength{\tabcolsep}{5pt}
\resizebox{\textwidth}{!}{
\begin{tabular}{lcccccc}
\toprule
\multirow{2}{*}{Method}
& \multicolumn{4}{c}{Rotation Attack: TPR / Acc $\uparrow$}
& \multirow{2}{*}{FID $\downarrow$}
& \multirow{2}{*}{CLIP $\uparrow$} \\
\cmidrule(lr){2-5}
& $\pm 5^\circ$
& $\pm 10^\circ$
& $\pm 30^\circ$
& $\pm 60^\circ$
& & \\
\midrule
Stable Diffusion v2.1
& -- 
& --
& --
& --
& $57.2958 \pm 0.9591$
& $0.3080 \pm 0.0015$ \\

\midrule
DwtDctSvd
& $0.3511 / 0.5215$
& $0.3394 / 0.5152$
& $0.3115 / 0.5069$
& $0.3153 / 0.5127$
& $56.7683 \pm 0.9571$
& $0.3122 \pm 0.0015$ \\

RivaGAN
& $0.9971 / 0.9974$
& $0.9928 / 0.9898$
& $0.6151 / 0.6358$
& $0.3992 / 0.4875$
& $56.8412 \pm 0.9864$
& $0.3108 \pm 0.0014$ \\

Stable Signature
& $0.9692 / 0.9620$
& $0.8827 / 0.7230$
& $0.6429 / 0.5833$
& $0.5193 / 0.5416$
& $57.5137 \pm 1.0388$
& $0.3068 \pm 0.0015$ \\

Tree-Ring
& $0.9560 / \text{--}$
& $0.6780 / \text{--}$
& $0.2700 / \text{--}$
& $0.5120 / \text{--}$
& $58.6148 \pm 0.9949$
& $0.3081 \pm 0.0014$ \\

RingID
& $0.8700 / \text{--}$
& $0.8840 / \text{--}$
& $0.7800 / \text{--}$
& $0.7340 / \text{--}$
& $58.1896\pm 0.8413$
& $0.3048 \pm 0.0013$ \\

\midrule

Gaussian Shading
& $0.0140 / 0.5658$
& $0.0000 / 0.5523$
& $0.0000 / 0.5040$
& $0.0000 / 0.5018$
& $58.0626 \pm 0.5199$
& $0.3088 \pm 0.0004$ \\

\rowcolor{phasebg}
\quad\textbf{+ AnchorMark}
& $\mathbf{0.9900} / \mathbf{0.9750}$
& $\mathbf{1.0000} / \mathbf{0.9822}$
& $\mathbf{1.0000} / \mathbf{0.9802}$
& $\mathbf{0.9960} / \mathbf{0.9689}$
& $59.2894 \pm 1.0701$
& $0.3082 \pm 0.0012$ \\

ShapeMark
& $0.5600 / 0.5098$
& $0.5820 / 0.5122$
& $0.4540 / 0.5001$
& $0.4200 / 0.4990$
& $56.4198 \pm 0.7918$
& $0.3096 \pm 0.0011$ \\

\rowcolor{phasebg}
\quad\textbf{+ AnchorMark}
& $\mathbf{1.0000} / \mathbf{1.0000}$
& $\mathbf{1.0000} / \mathbf{1.0000}$
& $\mathbf{1.0000} / \mathbf{0.9999}$
& $\mathbf{1.0000} / \mathbf{1.0000}$
& $57.4025 \pm 0.8211$
& $0.3088 \pm 0.0012$ \\

\bottomrule
\end{tabular}
}
\end{table*}

\section{Experiments}
\subsection{Experimental Settings}
Unless otherwise specified, experiments use Stable Diffusion v2.1 at \(512\times512\) resolution with a \(4\times64\times64\) initial latent. Images are generated with 50 denoising steps and a classifier-free guidance scale of 7.5. Watermark recovery uses 10-step DDIM inversion with a null prompt, assuming the original generation prompt is unavailable.
AnchorMark uses a \(16\times16\) central anchor over all four latent channels. The multi-frequency phase anchor contains two sine--cosine pairs with \(\mathcal{K}=\{1,3\}\), phase offsets \((\phi_1,\phi_2)=(0,0.7)\), and injection strength \(\lambda=0.2\). Phase-anchor-guided global registration uses a \(5^\circ\) coarse grid followed by \(1^\circ\) refinement within \(\pm3^\circ\). Payload-confidence-guided local refinement searches within \(\pm2^\circ\) using a \(0.5^\circ\) step. All experiments are conducted on an NVIDIA L40 GPU.

\subsection{Evaluation Metrics}
Following~\cite{yang2024gaussian}, we evaluate both watermark detection and tracing. Detection performance is measured by the true positive rate (TPR) at a fixed false positive rate of \(10^{-6}\), while tracing performance is measured by bit accuracy (Acc). Configurable multi-bit methods use a 256-bit payload; methods with fixed capacity follow their official settings and released weights. TPR and Acc are evaluated on 500 randomly sampled prompts from Stable-Diffusion-Prompts~\cite{stable_diffusion_prompts_gustavosta}. Image quality and text--image consistency are assessed using FID~\cite{heusel2017gans} and CLIP-Score~\cite{radford2021learning}, computed on 1,000 COCO2017 prompts~\cite{lin2014microsoft}.

\subsection{Baselines}
We compare AnchorMark with representative post-hoc methods, including DwtDctSvd~\cite{ingemar2008digital} and RivaGAN~\cite{zhang2019robust}; the fine-tuning-based Stable Signature~\cite{fernandez2023stable}; and inversion-based methods, including Tree-Ring~\cite{wen2023tree}, RingID~\cite{ci2024ringid}, Gaussian Shading~\cite{yang2024gaussian}, and ShapeMark~\cite{qian2026shapemark}. Detailed baseline configurations are provided in the appendix.

\subsection{Main Results: Comprehensive Rotation-Robustness Comparison}
We evaluate rotation robustness in both detection and tracing settings. 
Each \(\pm\theta^\circ\) setting evaluates both \(+\theta^\circ\) and \(-\theta^\circ\) rotation attacks, and the reported results are averaged over the two directions. 
Table~\ref{tab:main_results} reports TPR/Acc under different rotation magnitudes, together with FID and CLIP-Score for image quality and semantic fidelity. 
Rotation attacks severely degrade existing watermarking methods, especially inversion-based multi-bit schemes without synchronization: Gaussian Shading almost loses detection capability beyond small rotations, and both Gaussian Shading and ShapeMark keep tracing accuracy close to random guessing. 
By contrast, AnchorMark restores this coordinate frame through latent registration before decoding. 
With AnchorMark, Gaussian Shading achieves near-perfect detection and an average Acc of \(97.66\%\), while ShapeMark reaches almost perfect recovery across all tested rotations, including \(\pm60^\circ\). 
Meanwhile, the changes in FID and CLIP-Score remain limited compared with the corresponding base methods, indicating that AnchorMark provides accurate geometric correction while preserving visual quality and text-image semantic fidelity.

\subsection{Robustness under Combined Attacks}

While the main comparison focuses on isolated rotations, real editing pipelines often combine rotation with resizing, compression, noise, or filtering. 
These compound attacks are more challenging because they simultaneously disrupt spatial alignment and degrade the anchor or payload signal. 
We therefore evaluate AnchorMark under randomized combined attacks across SDv1.5, SDv2.1, SDv3.5 Medium, and FLUX.1-dev.
For each test image, we randomly compose one rotation with one additional image-level distortion. 
The rotation angle is sampled from \(\{15^\circ,30^\circ,45^\circ\}\), covering moderate to large geometric misalignment. 
The second distortion is sampled from five common post-processing operations: upscaling by \(1.25\times\), downscaling by \(0.9\times\), JPEG compression with quality factor \(75\), Gaussian noise with \(\sigma=0.01\), and median blur with kernel size \(5\). 
This randomized protocol avoids evaluating only manually selected attack pairs and better reflects practical editing scenarios where geometric and photometric degradations may appear jointly. 
We report the average TPR and Acc over all randomly sampled combined attacks.
Table~\ref{tab:cross_backbone_random_combined} summarizes the results. 
AnchorMark maintains stable detection and tracing performance across all evaluated backbones, showing that the proposed synchronization mechanism is not tied to a specific diffusion architecture or latent distribution. 
More importantly, the results demonstrate that AnchorMark remains effective when rotation is coupled with common image-level distortions. 
This indicates that the phase anchor provides a sufficiently robust synchronization cue, and that the two-stage correction-and-extraction strategy can still recover the payload after realistic compound transformations. 

\begin{table}[h]
\centering
\caption{
Average TPR / Acc under randomized combined attacks across text-to-image backbones. 
Each attack combines a rotation from \(\{15^\circ,30^\circ,45^\circ\}\) with a randomly selected resize, JPEG, Gaussian noise, or median blur distortion.
}
\label{tab:cross_backbone_random_combined}
\renewcommand{\arraystretch}{1.10}
\setlength{\tabcolsep}{4.2pt}
\resizebox{\columnwidth}{!}{
\begin{tabular}{lcc}
\toprule
\multirow{2}{*}{Backbone}
& \multicolumn{2}{c}{TPR / Acc \(\uparrow\)} \\
\cmidrule(lr){2-3}
& Gaussian Shading + AnchorMark
& ShapeMark + AnchorMark \\
\midrule
Stable Diffusion v1.5
& 0.9981 / 0.9642
& 1.0000 / 0.9768 \\

Stable Diffusion v2.1
& 0.9980 / 0.9598
& 1.0000 / 0.9722 \\

Stable Diffusion v3.5
& 0.9980 / 0.9527
& 1.0000 / 0.9641 \\

FLUX.1-DEV
& 0.9979 / 0.9454
& 0.9999 / 0.9678 \\

\bottomrule
\end{tabular}
}
\end{table}

\subsection{Ablation Study on Phase Anchor Design}

We ablate the phase-anchor template to verify whether the proposed multi-frequency design is necessary. 
All variants use the same anchor region, injection strength, and decoding strategy, and differ only in the injected synchronization pattern. 
We compare a statistics-matched random anchor, a low-frequency anchor using only the \(k=1\) harmonic, a high-frequency anchor using only the \(k=3\) harmonic, and the full multi-frequency anchor combining both components.

Table~\ref{tab:anchor_ablation} shows that the random anchor remains partially effective: since it is key-dependent and spatially fixed, it can still provide a certain matching signal after rotation. 
However, it lacks an analytic phase response, leading to less accurate angle estimation and lower tracing accuracy. 
The low-frequency anchor improves stability and achieves better accuracy than the random anchor, while the high-frequency anchor provides stronger angular sensitivity but is more affected by interpolation and inversion noise. 
The full multi-frequency anchor yields the best performance, reducing the angle estimation error to less than 0.5$^\circ$.
These results confirm that AnchorMark benefits from the structured combination of low-frequency robustness and high-frequency discrimination, rather than merely from inserting an additional latent pattern.

\begin{table}[h]
\centering
\caption{
Ablation study of the phase-anchor design under rotation attacks.
}
\label{tab:anchor_ablation}
\small
\renewcommand{\arraystretch}{1.08}
\begin{tabular*}{\columnwidth}{@{\extracolsep{\fill}}lcc@{}}
\toprule
\textbf{Anchor Design}
& \textbf{Angle Error} $(^\circ)$ $\downarrow$
& \textbf{TPR / Acc} $\uparrow$ \\
\midrule
\multicolumn{3}{@{}l}{\emph{Gaussian Shading + AnchorMark}} \\
\quad Random
& 0.4975 
& 0.9850 / 0.9388 \\
\quad Low-frequency only
& 0.4450 
& 0.9950 / 0.9529 \\
\quad High-frequency only
& 0.4050 
& 0.9900 / 0.9510 \\
\quad \textbf{Full multi-frequency}
& \textbf{0.3980} 
& \textbf{0.9980} / \textbf{0.9598} \\
\addlinespace[3pt]
\multicolumn{3}{@{}l}{\emph{ShapeMark + AnchorMark}} \\
\quad Random
& 0.6820
& \textbf{1.0000} / 0.9432 \\
\quad Low-frequency only
& 0.5710
& \textbf{1.0000} / 0.9581 \\
\quad High-frequency only
& 0.6070
& \textbf{1.0000} / 0.9488 \\
\quad \textbf{Full multi-frequency}
& \textbf{0.4929}
& \textbf{1.0000} / \textbf{0.9722} \\
\bottomrule
\end{tabular*}
\end{table}

\section{Conclusion}
We presented AnchorMark, a training-free robust inversion-based watermarking. 
AnchorMark embeds a multi-frequency phase anchor into the initial diffusion latent and restores the correct spatial coordinate frame through hierarchical rotation synchronization and payload recovery. 
Experiments show that AnchorMark significantly improves multi-bit tracing under rotation and combined attacks while preserving image quality and semantic fidelity. 
These results demonstrate that latent-space synchronization is an effective mechanism for robust inversion-based watermarking under geometric transformations.

\bibliography{aaai2027}

\end{document}